\documentclass[10pt]{article}

\pdfoutput=1
\usepackage{spconf}
\usepackage{amsmath}
\usepackage{amssymb}
\usepackage{amsfonts}
\usepackage{epsfig}
\usepackage{algorithm2e}
\usepackage{xcolor}

\def\reals{\ensuremath{\mathbb{R}}}
\def\naturals{\ensuremath{\mathbb{N}}}

\newcommand{\mat}[1]{\ensuremath{\mathbf{#1}}}
\newcommand{\st}{\ensuremath{\quad\mathrm{s.t.}\quad}}
\newcommand{\norm}[1]{\ensuremath{\left\|#1\right\|}}
\newcommand{\cost}[1]{\ensuremath{\ell_{#1}}}

\newcommand{\setdef}[1]{\ensuremath{\left\{#1\right\}}}

\newcommand{\opt}[1]{{#1}^*} \newcommand{\refeq}[1]{(\ref{#1})}

\def\sparsa{\textsc{s}pa\textsc{rsa}}

\def\dictm{\mat{D}}
\def\dictv{\mat{d}}

\def\datam{\mat{X}}
\def\datav{\mat{x}}

\def\coefm{\mat{A}}
\def\coefv{\mat{a}}
\def\coef{a}

\def\ndims{m}
\def\natoms{p}
\def\nsamples{n}
\def\group{g}
\def\groupset{\mathcal{G}}
\def\ngroups{{|\mathcal{G}|}}
\def\reg{\psi}
\def\sgn{\mathrm{sgn}}

\newcommand{\argmin}[1]{\underset{#1}{\operatorname{argmin}}}

\ninept

\begin{document}
\title{Collaborative Hierarchical Sparse Modeling}

\twoauthors{Pablo Sprechmann, Ignacio Ramirez and Guillermo
Sapiro}{University of Minnesota}{Yonina C. Eldar\thanks{IR and PS
contributed equally to this work.}}{Technion I. I. T.}

\maketitle
\begin{abstract}
Sparse modeling is a powerful framework for data analysis and
processing. Traditionally, encoding in this framework is done
by solving an $\ell_1$-regularized linear regression problem, usually
called {\it Lasso}. In this work we first combine
the sparsity-inducing property of the Lasso model, at the individual
feature level, with the block-sparsity property of the
{\it group Lasso} model, where sparse groups of features are jointly encoded, obtaining a sparsity pattern hierarchically
structured. This results in the {\it hierarchical
Lasso}, which shows important practical modeling advantages. We then
extend this approach to the collaborative case, where a set of simultaneously coded signals share the
same sparsity pattern at the higher (group) level but not necessarily at the lower
one. Signals then share the same active groups, or classes, but not necessarily
the same active set. This is very well suited for
applications such as source separation. An efficient optimization procedure, which guarantees convergence to the global optimum, is developed for these new models. The underlying presentation of the new framework and optimization approach is complemented with experimental examples and preliminary theoretical results.
\end{abstract}

\vspace{-10pt}
\section{Introduction and Motivation} 
In addition to being very attractive at the theoretical level, sparse signal modeling has been shown to lead to numerous state-of-the-art results in signal processing. The standard model assumes that a signal can be efficiently represented by a sparse linear combination of atoms from a given or learned dictionary. The selected atoms form what is usually referred to as the {\it active set}, whose cardinality is significantly smaller than the size of the dictionary  and the dimension of the signal. In recent years, it has been shown that adding structural constraints to this active set has value both at the level of representation robustness and at the level of signal interpretation (in particular when the active set indicates some physical properties of the signal), see  \cite{bach09}  and references therein. This leads to {\it group} or {\it structured} sparse coding, where instead of considering the atoms as singletons, the atoms are grouped, and a few groups are active at a time. An alternative way to add structure (and robustness) to the problem is to consider the simultaneous encoding of multiple signals, requesting that they all share the same active set. This is a natural collaborative filtering approach to sparse coding, see \cite{tropp06a} and references therein.

In this work we extend these models in a number of directions. First, we present a hierarchical sparse model, where not only a few (sparse) groups of atoms are active at a time, but also each group enjoys internal sparsity.\footnote{While we here consider only 2 levels of sparsity, the proposed framework is easily extended to multiple levels.} At the conceptual level, this means that the signal is represented by a few groups (classes), and inside each group only a few members are active at a time. A simple example of this is a piece of music (numerous applications in genomics), where only a few instruments are active at a time (each instrument is a group), and the actual music played by the instrument is efficiently represented by a few atoms of the sub-dictionary/group corresponding to it. Thereby, this proposed hierarchical sparse coding framework permits to efficiently perform source separation, where the individual sources (classes/groups) that generated the signal are identified at the same time as their efficient representation is reconstructed (the sparse code inside the group). An efficient optimization procedure is proposed to solve this hierarchical sparse coding framework.

Then, we go a step beyond this. Imagine now that we have multiple recordings of the same two instruments (or different time windows of the same recording), each time playing different songs. Then, if we apply this new hierarchical sparse coding approach collaboratively, we expect that the different recordings will share the same groups (since they are of the same instruments), but each will have its unique sparsity pattern inside the group (since each recording is a different melody). We propose a collaborative hierarchical sparse coding framework addressing exactly this.\footnote{Note that different recordings can also have different instruments, so some of them will share the same groups while not necessarily all of them will be exactly the same.}  An efficient optimization procedure for this case is derived as well.

In the remainder of this paper, we introduce these new models and their corresponding optimization, present examples illustrating them, and provide possible directions of research opened by these new frameworks, including some theoretical ones.

\vspace{-10pt}

\section{Collaborative Hierarchical Coding}

\subsection{Background: Lasso and group Lasso}
\label{sec:intro}


Assume we have a set of data samples $\datav_j \in \reals^{\ndims},
j=1,\ldots,\nsamples$, and a dictionary of $\natoms$ atoms, assembled
as a matrix $\dictm \in \reals^{\ndims{\times}\natoms}$, $\dictm=[\dictv_1
\dictv_2 \ldots \dictv_\natoms]$. Each sample $\datav_j$ can be written as
$\datav_j=\dictm\coefv_j+\epsilon,\,\coefv_j \in \reals^{\natoms},\,\epsilon \in
\reals^{\ndims}$, that is, as a linear combination of the atoms in the
dictionary $\dictm$ plus some perturbation $\epsilon$. The basic
underlying assumption in sparse coding is that, for all or most $j$, the
optimal reconstruction $\coefv_j$ has only a few nonzero elements.  Formally,
if we define the cost $\cost{0}$ as the pseudo-norm counting the number of
nonzero elements of $\coefv_j$, $\norm{\coefv_j}_0:=|\{k:\coef_{kj} \neq 0\}|$,
we expect that $\norm{\coefv_j}_0 \ll \natoms$ for all or most $j$.
The $\ell_0$ optimization is non-convex and known to be \textsc{NP}-hard, so a convex
approximation to it is considered instead, which uses the $\cost{1}$ norm
cost,
\begin{equation}
\min_\coefv \norm{\coefv}_1 \st \norm{\datav_j-\dictm\coefv}_2^2 \leq \epsilon.
\end{equation}
The above approximation is known as the Lasso \cite{tibshirani96}. A popular
variant is to use the unconstrained version
\begin{equation}
\min_\coefv \frac{1}{2}\norm{\datav_j-\dictm\coefv}_2^2 + \lambda\norm{\coefv}_1,
\label{eq:lasso}
\end{equation}
where $\lambda$ is a parameter usually found by cross-validation.

%
%



The $\norm{\cdot}_1$ regularizer  induces sparsity in the solution $\coefv_j$.
This is desirable not only from a regularization point of view, but also from a model selection point, where one
wants to identify the relevant features or factors (atoms) that conform each sample $\datav_j$.
In many situations, however, one wants to represent the relevant factors not
as single atoms but as groups of atoms. Given a dictionary of $\natoms$ atoms, we define groups
through their indexes, $\group \subseteq \{1,\ldots,\natoms\}$.  Given a
group $\group$, we define the subset of atoms of $\dictm$ belonging to it as $\dictm_\group$, and the corresponding set of linear reconstruction
coefficients as $\coefv_\group$.  Define
$\groupset=\{\group_1,\ldots,\group_\ngroups \}$ to be a partition of
$\{1,\ldots,\natoms\}$. The group Lasso problem was introduced in \cite{yuan06} as
\begin{equation}
\min_\coefv \frac{1}{2}\norm{\datav_j-\dictm\coefv}_2^2 + \lambda\reg_{\groupset}(\coefv),
\label{eq:group-lasso}
\end{equation}
where $\reg_\groupset$ is the group Lasso regularizer defined in
terms of $\groupset$ as $\reg_\groupset(\coefv) = \sum_{\group \in \groupset}{\norm{\coefv_\group}_2}$.
Note that $\reg_{\groupset}$ can be seen as an $\cost{1}$ on Euclidean norms
of the vectors formed by coefficients belonging to the same group
$\coefv_\group$. This is a generalization of the $\cost{1}$ regularizer, as
the latter arises from the special case
$\groupset=\setdef{1,2,\ldots,\natoms}$, and, as such its effect
on the groups of $\coefv$ is also a natural generalization of the one
obtained with the Lasso: it ``turns on'' or ``off'' atoms in groups.

\subsection{The Hierarchical Lasso}

The group Lasso trades sparsity at the single-coefficient level with sparsity
at a group level, while, inside each group, the solution is dense (actually
it reduces to a least squares within the group). As we are interested in
maintaining the sparsity at the coefficient level, we simply re-introduce the
$\cost{1}$ regularizer together with the group regularizer, leading to the
proposed {\it Hierarchical Lasso (HiLasso)} model,\footnote{While preparing the camera ready version of this work
we leaned of a simultaneously developed paper, \cite{Stanford}, that also proposed this model, with a different optimization approach.
The collaborative framework presented next is not developed in \cite{Stanford}. See also \cite{peng}.}
\begin{equation}
\min_\coefv \frac{1}{2}\norm{\datav_j-\dictm\coefv}_2^2 +
\lambda_2\reg_{\groupset}(\coefv) + \lambda_1\norm{\coefv}_1.
\label{eq:hilasso}
\end{equation}
We refer to this regularizer as the $\ell_2 + \ell_1$.\footnote{We can similarly define a hierarchical
sparsity model based on $\ell_0$.}
In Section~\ref{sec:opt} we propose an efficient optimization for \eqref{eq:hilasso},
while in Section~\ref{sec:results} we experimentally show the virtues of this model.

\subsection{Collaborative Hierarchical Lasso}
\label{sec:background:collaborative-lasso}

In numerous applications, one expects that certain collections of samples
$\datav_j$ share the same active components from the dictionary, that is,
that the indexes of the nonzero coefficients in $\coefv_j$ are the same
for all the samples in the collection. Imposing such dependency in the
$\cost{1}$ regularized regression problem gives rise to the so called
collaborative (also called ``multitask'' or ``simultaneous'') sparse
coding problem \cite{tropp06a,wright04}.

More specifically, if we consider the matrix of coefficients
$\coefm=[\coefv_1,\ldots,\coefv_\nsamples]$ associated to the reconstruction of
the samples $\datam=[\datav_1,\ldots,\datav_\nsamples]$, the collaborative sparse coding
model is given by
\begin{equation}
\min_{\coef} \frac{1}{2}\norm{\datam - \dictm\coefm}_F^2 +
 \lambda \sum_{k=1}^{\natoms}{\norm{\coefv^k}_2},
\label{eq:collaborative-lasso}
\end{equation}
where $\coefv^k$ is the $k$-th row of $\coefm$, that is, the vector of the
$\nsamples$ different values that the coefficient associated to the $k$-th
atom takes for each sample $j$.  If we now extend this idea to the group
Lasso, we obtain a collaborative group Lasso formulation,
\begin{equation}
\min_{\coef} \frac{1}{2}\norm{\datam- \dictm\coefm}_F^2 +
 \lambda\reg_{\groupset}(\coefm),
\label{eq:collaborative-group-lasso}
\end{equation}
where the regularizer $\reg_{\groupset}$ for a matrix is defined as
%
$\reg_\groupset(\coefm) = \sum_{\group \in \groupset}{\norm{\coefm_\group}_F}$,
%
being $\coefm_\group$ the submatix formed by all the rows belonging to group $\group$.\footnote{While the introduced collaborative HiLasso model is more general, we consider the separable
case for the optimization here developed.} We chose
this notations since this regularizer is the natural extension of the regularizer in \refeq{eq:group-lasso} for the collaborative case.

To the best of our knowledge, this combination has not yet been investigated in the
literature.  In this paper we are moving one step forward and treat
this together with the hierarchical extension.
The combined model that we propose for this problem ({\it C-HiLasso}) can be written as follows
\begin{equation}
\min_{\coefm} \frac12 \norm{\datam - \dictm\coefm}_F^2
+   \lambda_2 \reg_{\groupset}(\coefm) +\sum_{j=1}^{\nsamples}  \lambda_1\norm{\coefv_j}_1.
\label{eq:collaborative-hilasso}
\end{equation}
The collaborative group Lasso is a particular case of our model when $\lambda_1$ is zero. On the other hand, one can obtain independent
Lasso for each $\mat{x}_i$ by setting $\lambda_2$ to zero. This new formulation is particularly well suited when the
vectors have missing components. In this case combining the information from all the samples is very important
in order to lead to a correct representation and model (group) selection. This can be done by slightly changing the data term in \eqref{eq:collaborative-group-lasso}.
For each data vector $\datav_j$ one computes the reconstruction error using only the observed elements. Note that the missing components do not affect the other terms of the equation.


\section{Optimization}
\label{sec:opt}

\subsection{Single-signal problem: HiLasso}
\label{sec:opt.single}

In the last decade, optimization of problems of the form of \eqref{eq:lasso} and \eqref{eq:group-lasso} have been deeply studied
and there exist very efficient algorithms for solving them.
Recently, Wright et. al \cite{SpaRSA} proposed a framework, \sparsa, for solving the general problem
\begin{equation}
\min_\coefv f(\coefv) + \lambda\reg(\coefv),
\label{eq:general.problem}
\end{equation}
 under reasonable assumptions. To guarantee convergence $f$ needs to be a smooth and convex function while $\reg$ only needs to be finite in $\reals^n$.
When the regularizer, $\reg$, is group separable, the optimization can be subdivided into smaller problems, one per group. The framework becomes powerful when these subproblems  can be solved efficiently.
This is the case of the Lasso and group Lasso settings but is not immediate when the regularizer is  the proposed $\ell_1 + \ell_2$ norm.
In this work we combine the \sparsa\ with the Alternating Direction Method of Multipliers \cite{bertsekas89} (\textsc{admom}),
to efficiently solve the HiLasso problem.

The \sparsa\ algorithm generates a sequence of iterates $\{ \mat{x}^t \}_{t \in  \naturals}$ that, under certain conditions,
converges to the solution of \eqref{eq:general.problem}.  At each iteration, $\mat{x}^{t+1}$ is obtained solving
\begin{equation}
\min_\mat{z} (\mat{z}-\mat{x}^t)^T \nabla f(\mat{x}^t) + \frac{\alpha^t}{2} \norm{ \mat{z} - \mat{x}^t}^2_2 + \lambda \reg(\mat{z}),
\label{eq:sparsa-subproblem}
\end{equation}
\noindent for some sequence of parameters $\{ \alpha^t \}_{t \in \naturals}$ with $\alpha^t \in \reals^{+}$. The conditions for which the algorithm converges
depend on the choice of  $\alpha^t $, see \cite{SpaRSA} for details.

It is easy to show that (\ref{eq:sparsa-subproblem}) is equivalent to
\begin{equation}
\min_\mat{z} \frac12 \norm{\mat{z} - \mat{u}^t}^2_2 + \frac{\lambda}{\alpha^t} \reg(\mat{z}),
\label{eq:sparsa-subproblemEq}
\end{equation}
\noindent where $ \mat{u}^t= \mat{x}^t - \frac{1}{\alpha^t}\nabla f(\mat{x}^t).$
In this new formulation, it is clear that the first term in the cost
function can be separated element-wise. Thus when the regularization function
$\reg(\mat{z})$ is group separable, so is the overall optimization, and one can solve
(\ref{eq:sparsa-subproblemEq}) independently for each group,
\begin{equation*}
\min_{\mat{z}_{\group}} \frac12 \norm{ \mat{z}_{\group} - \mat{u}^{t}_{\group}}^2_2 + \frac{\lambda}{\alpha^t} \reg_{\group}(\mat{z}_{\group}),
\end{equation*}
which in the case of HiLasso, this becomes,
\begin{equation}
\min_{\mat{b}\in\reals^{|\group|}}
\frac{1}{2}\norm{\mat{b}-\mat{w}}_2^2 +
\frac{\lambda_2}{\alpha^t} \norm{\mat{b}}_2 + \frac{\lambda_1}{\alpha^t}\norm{\mat{b}}_1,
\label{eq:sparsa:sub}
\end{equation}
\noindent where $\mat{w} = \mat{u}^t_{\group}$ and $\mat{u}^t
=\mat{a}^t - \frac{1}{\alpha^t} \dictm^T( \dictm \mat{a}^t -\datav) $.
This is a \textsc{socp} for which one could use generic solvers.
However, this subproblem needs to be solved many times within the \sparsa\  iterations, so it is crucial to solve it efficiently.
For this we use the \textsc{admom} method \cite{bertsekas89}.
The idea is to solve the artificially constrained equivalent problem,
\begin{equation*}
\min_b
\frac12 \norm{\mat{b}-\mat{w}}_2^2 +
\tilde{\lambda}_2\norm{\mat{\beta}}_2  +
\tilde{\lambda}_1\norm{\mat{b}}_1, \, \st \mat{b}=\beta,
\end{equation*}
where $\tilde{\lambda}_i = \lambda_i/\alpha^t$. The algorithm generates a set of iterates $\{ \mat{b}^t, \beta^t, \mat{p}^t \}_{t \in \naturals^{+}}$ which converges to the minimum of the Augmented Lagrangian of the problem
\begin{align*}
L_c(\mat{b},\beta,\mat{p})  = &
\frac12 \norm{\mat{b}-\mat{w}}_2^2 +
\tilde{\lambda}_2\norm{\mat{b}}_2 +
\tilde{\lambda}_1\norm{\mat{b}}_1 \\
&+\mat{p}^T(\mat{b} - \mat{\beta})
+ \frac{c}{2}\norm{\mat{b} - \mat{\beta}}_2^2,
\end{align*}
where the elements of $\mat{p}$ are the so called Lagrangian multipliers, and $c$ is a fixed constant.
At each iteration, the variables $\mat{b}$ and $\beta$ are updated, one at a time, by minimizing the Augmented Lagrangian while letting the remaining fixed:
\begin{align}
\mat{b}^{t+1} = &\argmin{\mat{b}}
\frac12 \norm{\mat{b}-\mat{w}}_2^2 +
\tilde{\lambda}_1\norm{\mat{b}}_1 +
\mat{b}^T\mat{p} \quad \nonumber \\
 & + \frac{c}{2}\norm{\mat{b} - \mat{\beta}}_2^2,\label{eq:actBeta}\\
\mat{\beta}^{t+1} =& \argmin{\beta} \,\,
\tilde{\lambda}_2\norm{\mat{\beta}}_2 -
\mat{\beta}^T \mat{p}+ \frac{c}{2}\norm{\mat{b}^{t+1} - \mat{\beta}}_2^2,\nonumber\\
\mat{p}^{t+1} = & \mat{p}+ c(\mat{b}^{t+1}-\beta^{t+1}).\nonumber
\end{align}
%
For convenience in the notation we omitted the super-indexes for the iterates at step $t$, just explicitly indexing them at step $t+1$. The update for $\mat{b}$ is separable into scalar subproblems on
the coordinates of $\mat{b}$. The optimality conditions on
the subgradient of each of this scalar problems leads to a simple variant of the well
known soft-thresholding operator, $\mathcal{S}(w_i,\lambda)=\sgn(w_i)\max\setdef{0,|w_i|-\lambda}$. For convenience, we use the notation $\mathcal{S}(\mat{w},\lambda)$
to denote the vector obtained when applying the soft-thresholding operator (with parameter $\lambda$) to each element of $\mat{w}$.
On the other hand, the update for $\beta$ is not separable into scalar subproblems. However its optimality condition is given by
$\beta' + \tilde{\lambda}_2\partial\norm{\beta'}_2 - \mat{b}' \ni \mat{0}$,
which is exactly the one leading to the vector shrinkage operator, $\mathcal{S}_v$,
described in \cite{yuan06} for the group Lasso (actually much simpler, since
there is no matrix multiplication involved):
$$\mathcal{S}_v(\mat{b},\tilde{\lambda}_2) = \left[1 - \frac{\tilde{\lambda}_2}{\norm{\mat{b}}_2} \right]_{+}\!\mat{b}. $$
Then both updates can be written in closed form and computed very efficiently:
\begin{align*}
\mat{b} = \frac{1}{c\!+\!1} \mathcal{S}(\mat{w}\! +\! c\beta\! - \mat{p},\tilde{\lambda}_1), \, \,\beta  = \frac{1}{c}\,\mathcal{S}_v(\mat{p}\!+c\mat{b},\tilde{\lambda}_2).
\end{align*}
The algorithm is very robust and converges in very few iterations to its optimum, thereby obtaining a very efficient approach to solve
the subproblem \eqref{eq:sparsa:sub}. The \sparsa\ framework then becomes a very interesting approach for the proposed HiLasso.
The complete algorithm is summarized in Algorithm~\ref{alg1}. An additional speed up is obtained by bypassing \textsc{admom} when a whole group
is not active. From the optimality conditions of  \eqref{eq:sparsa:sub} it follows that, if $\mat{0}$ is a solution when $\lambda_1=0$ (standard group Lasso), it is also a solution in the general case. This can be simply checked by evaluating $\mathcal{S}_v(\,\mat{w}\,,\tilde{\lambda}_2)>\mat{0}$.

\begin{algorithm}[t]
{\footnotesize
\caption{{\footnotesize HiLasso optimization algorithm.}}
\label{alg1}
\KwResult{The optimal point $\opt{\mat{x}}$}
Set $t := 0$\;
Choose a factor $\eta>1$ and constants $c>0$ and $0<\alpha_{\textrm{min}}<\alpha_{\textrm{max}}$\;
Choose an initial  $\mat{x}(0)=(\mat{x}_{1},\mat{x}_{2},\ldots,\mat{x}_{\ngroups})$\;
\While{stopping criterion is not satisfied}{
Choose $\alpha^t\in [\alpha_{\textrm{min}},\alpha_{\textrm{max}}]$\;
Set $\mat{u}^t \leftarrow  \mat{x}^t - \frac{1}{\alpha^t}\nabla f(\mat{x}^t)$\;
\While{stopping criterion is not satisfied}{
\% Here we use the group separability of \eqref{eq:sparsa-subproblemEq} and \\
\% solve \eqref{eq:sparsa:sub} for each group\;
\For{$i=1$ to $\ngroups$}{
\eIf{$\mathcal{S}_v(\mat{w},\tilde{\lambda}_2)>\mat{0}$}{
Set $r := 0$\;
Choose an initial  $\mat{p}^0,\beta^0,\mat{b}^0$\;
\While{stopping criterion is not satisfied}{
 $\mat{b}^{r+1} = \frac{1}{c+1} \mathcal{S}(\mat{u^t_i} + c\beta^{r}- \mat{p^{r}},\tilde{\lambda}_1)$\;
 $\beta^{r+1} = \frac{1}{c}\mathcal{S}_v(\mat{p}^{r}+c\,\mat{b}^{r+1},\tilde{\lambda}_2)$\;
 $\mat{p}^{r+1} = \mat{p}^{r} + c(\mat{b}^{r+1}-\beta^{r+1})$\;
 Set $r \leftarrow r + 1$ \;
}
Set $\mat{x}^{t+1}_{\group} := \mat{b}^{r+1}$ \;
}{Set $\mat{x}^{t+1}_{\group} := \mat{0}$\;}
}
Set $\alpha^t \leftarrow \eta\alpha^t$\;
}
Set $t \leftarrow t + 1$ \;
}}
\end{algorithm}


\vspace{-10pt}

\subsection{Optimization of the Collaborative HiLasso}
\label{sec:collaborative-solution}

\begin{figure}
\begin{center}
\includegraphics[width=0.45\textwidth]{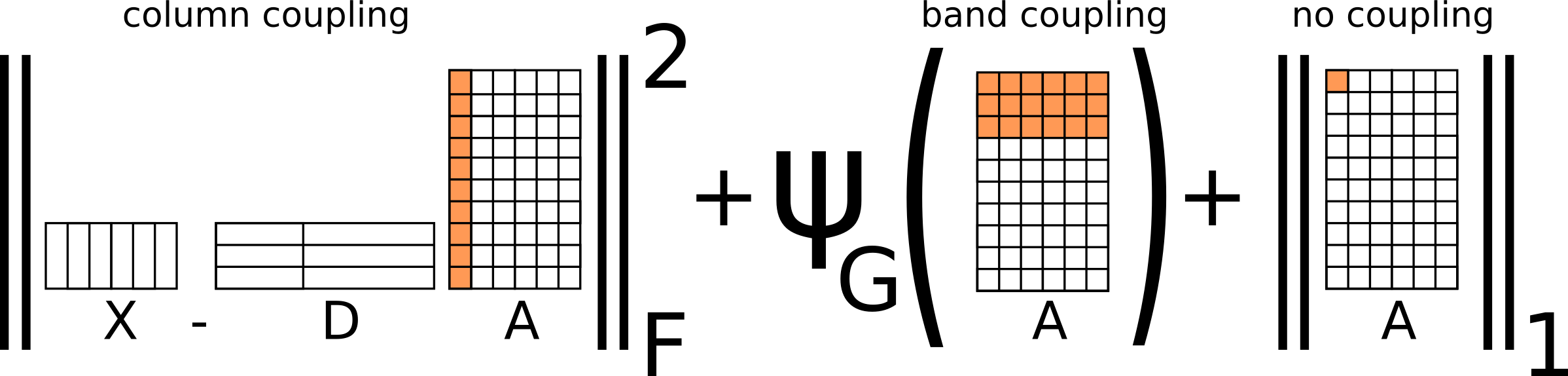}
\caption{\label{fig:structure} {\footnotesize Structure of the problem in terms of coupling.}}
\vspace{-.2in}
\end{center}
\end{figure}

\noindent We now propose an optimization algorithm to
efficiently solve the collaborative HiLasso. The main idea is to use
\textsc{admom} to divide the overall problem into two subproblems: one that breaks the
multi-signal problem into $\nsamples$ single-signal $\ell_1$ regressions, and another that
treats the multi-signal case as a single group Lasso-like problem. In this way we take advantage
of the separability of each term as shown in Figure~\ref{fig:structure}.
We define a constrained optimization problem,
\begin{align*}
\min \frac{1}{2}\norm{\datam\!-\!\dictm \coefm }_F^2\!  + \lambda_1 \!\sum_j \!\norm{\mat{a}_j}_1\!+\! \lambda_2 \reg_{\groupset}(\mat{B}) \,\,\, \mathrm{s.t.} \,\coefm \!\!= \!\!\mat{B}.
\end{align*}
The \textsc{admom} iterations are given by (we omitted the super-index for variables at iteration $t$ for notation convenience).
\begin{align}
\mat{\coefm}^{t+1} = & \argmin{\coefm}   \frac{1}{2}\norm{\datam-\dictm \coefm }_F^2 + \lambda_1 \sum_j \norm{\mat{a}_j}_1 + \nonumber\\
& \mathrm{Tr}(  \coefm^T\mat{P}^{t+1}) + \frac{c}{2}\norm{\mat{B} -\coefm }_F^2, \label{eq:col:problem1}\\
\mat{B}^{t+1}     = & \argmin{\mat{B}} \frac{c}{2}\norm{\mat{B} -\coefm^{t+1} }_F^2 + \mathrm{Tr}(\mat{B}^T \mat{P}^{t+1} ) \nonumber\\
&+ \lambda_2 \reg_{\groupset}(\mat{B}), \label{eq:col:problem2}\\
\mat{P}^{t+1}   =&  \mat{P} + c(\coefm-\mat{B}).\nonumber
\end{align}
\noindent \textbf{Solving for $\mat{A}^{t+1}$:} Problem \refeq{eq:col:problem1} can be separated into $n$ single-signal
subproblems by updating one column of the matrix $\coefm$ at a time,
\begin{equation*}
\min_{\mat{a}_j} \frac{1}{2}\norm{\datav\!-\dictm \mat{a}_j}_2^2 +
            \mat{p}_j^T \mat{a}_j +  \frac{c}{2}\norm{\mat{a}_j\!-\mat{b}}_2^2 +
            \lambda_1\norm{\mat{a}_j}_1.
\end{equation*}
This problem can be solved using the \sparsa\ framework. The idea is to consider the first three terms of the cost
as $f(\cdot)$ in Equation \eqref{eq:general.problem}. The associated computational cost is equivalent to the one of the Lasso, since the regularizer is the standard $\ell_1$ norm.

\noindent \textbf{Solving for $\mat{B}^{t+1}$:} The problem given by \refeq{eq:col:problem2} is group separable,
as a direct consequence of the separability of $\reg_{\groupset}$. Thus, we need to solve
$|\groupset|$ optimization problems of the form,
\begin{equation*}
\min_{\mat{B}_{\group}}  \frac{c}{2} \norm{\mat{B}_{\group} -\coefm_{\group}^{t+1} }_F^2 + \mathrm{Tr}( \mat{P}_{\group}^{t+1}\mat{B}_{\group}^T ) + \lambda_2 \norm{ \mat{B}_{\group} }_F,
\end{equation*}
\noindent where $\mat{A}_{\group}$, $\mat{B}_{\group}$ and
$\mat{P}_{\group}$ are the $|\group| \times n$ sub-matrices of
$\mat{A}$, $\mat{B}$ and $\mat{P}$ associated with the group $\group$
respectively.  We express them as column vectors (each with $|\group|{\times}n$
components) by concatenating their columns, obtaining $\mat{b}_\group, \beta_\group$ and
$\mat{p}_\group$ respectively,
and rewrite the optimization problem in vectorial form as
\begin{equation}
\min_\mat{b} \lambda_2\norm{\mat{b}}_2 - \mat{p}_{\group}^T\mat{b}+
                 \frac{c}{2}\norm{\mat{a}_{\group}^{t+1}-\mat{b}}_2^2.
\end{equation}
\noindent This problem is identical to \eqref{eq:actBeta} and can be reduced to a group Lasso
problem by simply changing variables and thus, it is solved using vectorial thresholding.



\section{Experimental results}
\label{sec:results}

We start by comparing our model with the standard Lasso and group Lasso using synthetic data.
We created $\ngroups$ dictionaries, $\mat{D}_i$, with 64 atoms of dimension 64, with i.i.d. Gaussian entries. The columns were
normalized to have unit $\ell_2$ norm. Then we randomly chose two groups to be active at each time (on all the signals).
Sets of $N=200$ testing signals were generated, one per active group, as linear combinations of $k \ll 64$
elements of the dictionaries, $\mat{x}_j^i = \mat{D}_i \coefv_j^i$. These signals were also normalized. The mixtures were created by summing
these signals and (eventually) adding gaussian noise of standard deviation $\sigma$. The generated testing signals
have a hierarchical sparsity structure and while they share groups, they do not necessarily share the sparsity pattern inside the groups.

We built a single dictionary by concatenating the sub-dictionaries, $\mat{D} = [\mat{D}_1,\ldots,\mat{D}_\ngroups]$, and use
it to solve the Lasso, group Lasso, HiLasso and C-HiLasso problems.
%
%
Table~\ref{tab:multi-signal-mse} summarizes the Mean Square Error (\textsc{mse}) and Hamming distance of the recovered coefficient vectors. We observe that our model is able to exploit the
hierarchical structure of the data as well as the collaborative structure.
From a modeling point of view, we observe that the group Lasso selects in general the correct blocks but it does not give a sparse solution within them. On
the other hand, Lasso gives a solution that has nonzero elements
belonging to groups that were not active in the original signal, leading to a wrong model selection. HiLasso gives a sparse solution that picks atoms form the correct groups but still
presents some minor mistakes. For the collaborative case, in all the tested cases, no coefficients were selected outside the correct active groups and the recovered coefficients are consistently the best ones.
This robustness comes from the fact that the active groups are collaboratively found using the information present in all the signals.
\begin{table}
\begin{center}
{\footnotesize
\begin{tabular}{|c|c|c|c|c|}\hline
$\sigma$     &Lasso  &Glasso&HiLasso & C-HiLasso     \\\hline  \hline
0.1               & 41.7/   22.0                              & 117.3 /              361.6          & 33.0       /  19.8                      &\textbf{16.3} / \textbf{13.3} \\
0.2               & 56.4 / 21.6                               & 118.2 / 378.3                         &    39.9     /  22.7                     &\textbf{24.9} /\textbf{17.1}  \\
0.4               & 96.5 / 22.7                               &  137.8/ 340.3                         &    65.6     /  \textbf{19.5}                    &\textbf{59.5} /27.4  \\ \hline
$k$     &Lasso &Glasso &HiLasso &C-HiLasso     \\\hline
8         & 38.8 /  22.0                            & 118.4 / 318.2                        & 27.2        /  19.5                      &\textbf{9.6} / \textbf{16.2}\\
12      & 120.0 / 36.2                           & 116.6 /350.4                        & 70.4       / \textbf{26.5}                         &\textbf{41.3} / 29.1\\
16      & 164.1 / 43.9                           & 109.3 / 338.6                         & 110.0    / \textbf{32.2}                        &\textbf{55.1} /  35.0\\ \hline
$\ngroups$     &Lasso  &Glasso &HiLasso &C-HiLasso     \\\hline
4      & 108.0 /  27.8                                          & 191.6 /221.7                         & 100.9      /     \textbf{29.8}                      &\textbf{74.2} / 30.2\\
8       & 120.0 /  36.2                                         & 116.6 /350.4                    & 70.4        / \textbf{26.5}                         &\textbf{41.3} / 29.1\\
\hline
\end{tabular}
}
\end{center}\vspace{-15pt}
\caption{\label{tab:multi-signal-mse} {\footnotesize Active sets \textsc{mse}  (we show them multiplied by $10^3$) and Hamming distance (MSE / Hamming) for the tested methods.
 In the first case we vary the noise level while we keep
  $\ngroups = 8$ and $k=8$ fixed. In the two other tables the signals are noise free and we first set $\ngroups = 8$ while changing $k$,
  and then set $k=12$ while changing the number of groups. For each method the regularization parameters were the ones for which the best results where obtained. }}
  \vspace{-.25in}
\end{table}
%
We consider the USPS digits dataset that has been shown to be well represented in the sparse modeling framework  \cite{CVPR}.
Here the signals are vectors containing the unwrapped gray intensities of $16\times 16$ images. We chose two digits and summed them up
to create a mixture image. We created 200 random mixture images and then analyzed them with the different methods.
In this case there is no ground truth active set, and we used as a measure of performance the separating error defined as
$\frac{1}{NR} \sum_{i=1}^{R} \sum_{j=1}^{N} \norm{\mat{x}_j^i - \hat{\mat{x}}_j^i}_2^2$,
where $\mat{x}_j^i$ is the component corresponding to source $i$ in the signal $j$, and $\hat{\mat{x}}_j^i$ is the recovered one.

\begin{figure}[t]
\begin{center}
\begin{tabular}{|c|cccc|}\hline
Digits        &Lasso  &Glasso &HiLasso & C-HiLasso     \\\hline  \hline
3+5           & 74.1 & 80.1 & 68.6 & \textbf{63.4} \\
3+5+$n$  & 87.9 & 95.4 & 92.9 & \textbf{77.3}\\
2+7           & 61.1 & 60.8 & 58.7 & \textbf{42.6}\\
2+7+$n$  & 75.4 & 65.2 & 64.7 & \textbf{53.7} \\\hline
\end{tabular}\\[5pt]
\includegraphics[width=0.426\textwidth]{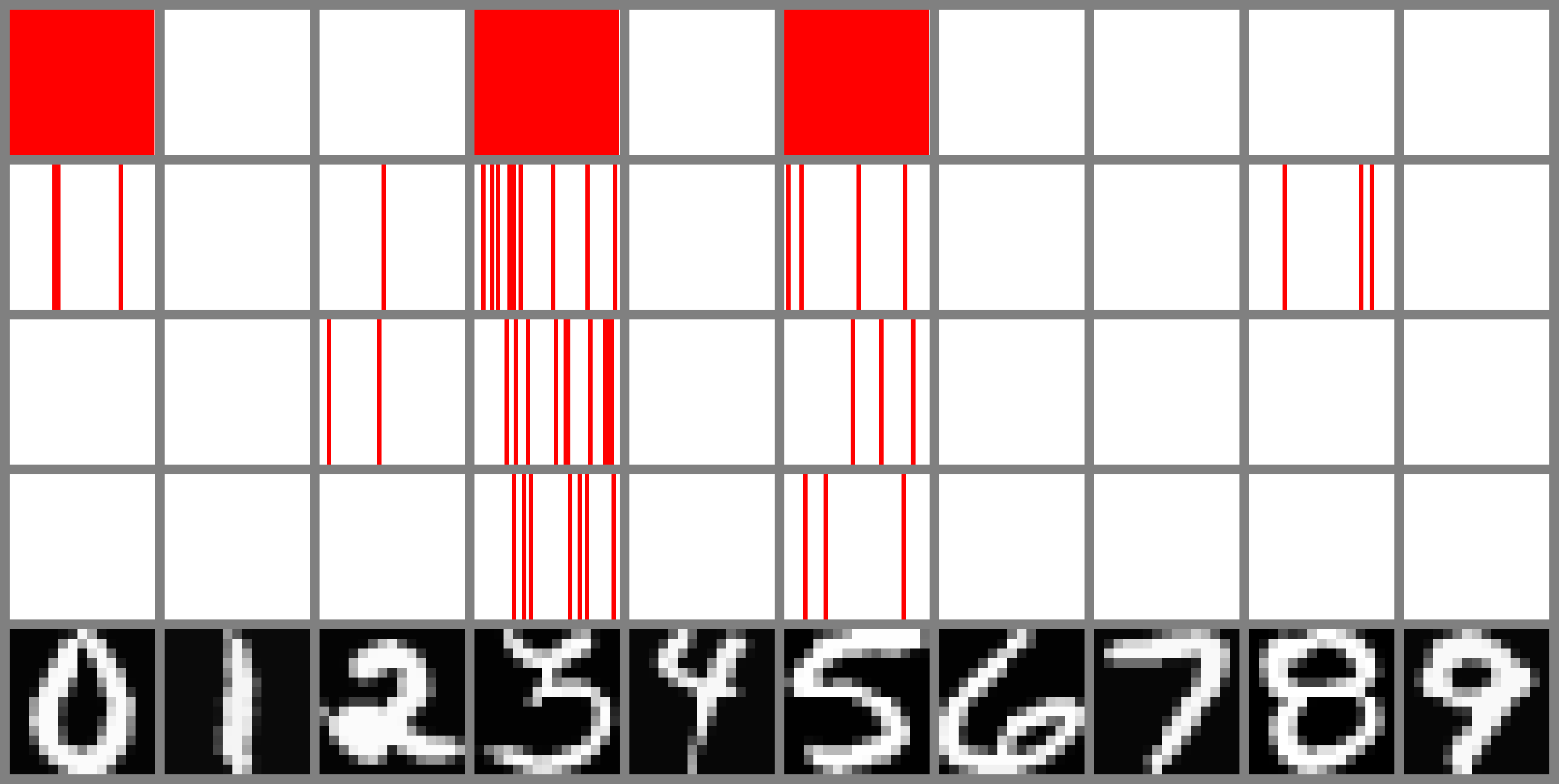}\\
\end{center}\vspace{-15pt}
\caption{\label{fig:digits}. {\footnotesize (Top) The table shows the separating errors (we show them multiplied by $10^3$) for the digits dataset. We show the results for
separating digits 3 and 5, and 2 and 7, with and without additive noise of standard deviation $\sigma=0.1$. We used sets of 200 copies. (Bottom) Active sets recovered
for the  group Lasso, Lasso, HiLasso and C-HiLasso for a given example. Each block corresponds to the coefficients associated with the
digits displayed bellow. The active coefficients are displayed in read. Only C-HiLasso manages to perfectly recover the correct models (with the lowest separating error), while HiLasso
performs very well also.} }
\vspace{-.2in}
\end{figure}

Using the usual training-testing split for USPS we first learned a dictionary for each digit. We then created a single dictionary
by concatenating them. In Figure~\ref{fig:digits} we show the separation error obtained in different situations. As in the synthetic case,
only the collaborative method was able to successfully detect the true active sources. We show in Figure~\ref{fig:digits} some examples of the recovered
active sets for each method.

\begin{figure}[t]
\begin{center}
\includegraphics[width=0.11\textwidth]{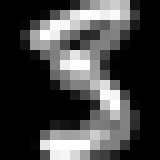}
\includegraphics[width=0.11\textwidth]{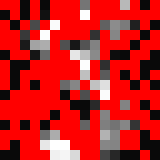}
\includegraphics[width=0.11\textwidth]{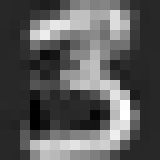}
\includegraphics[width=0.11\textwidth]{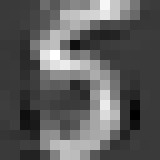}\\[1pt]
\includegraphics[width=0.11\textwidth]{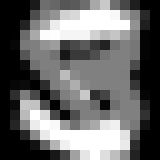}
\includegraphics[width=0.11\textwidth]{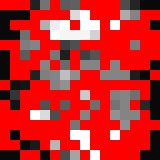}
\includegraphics[width=0.11\textwidth]{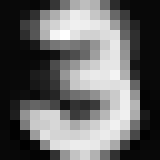}
\includegraphics[width=0.11\textwidth]{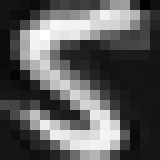}\\[2pt]
\includegraphics[width=0.23\textwidth,height=0.23\textwidth]{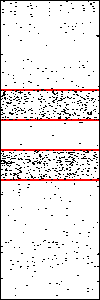}
\includegraphics[width=0.23\textwidth,height=0.23\textwidth]{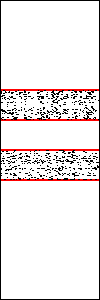}
\end{center}\vspace{-15pt}
\caption{\label{fig:missing}. {\footnotesize (Top) We show two examples (one per row) of the recovered digits from a mixture with 60\%
of missing components. We first show the original mixture image, then the image with the missing pixels highlighted in red, and finally the digits recovered. (Bottom)
Here we show a comparision of the active sets recovered using the Lasso (left) and the C-HiLasso (right) methods. The active sets for the set of signals (as shown in Figure~\ref{fig:digits}) are placed as columns. The coefficients corresponding to digits 3 and 5 fall inside the area delimited by the red horizontal lines. While C-HiLasso recovers  the correct sources in all the cases, the Lasso method makes several mistakes. } }
\vspace{-.2in}
\end{figure}

\begin{figure}[t]
\begin{center}
\includegraphics[width=0.48\textwidth]{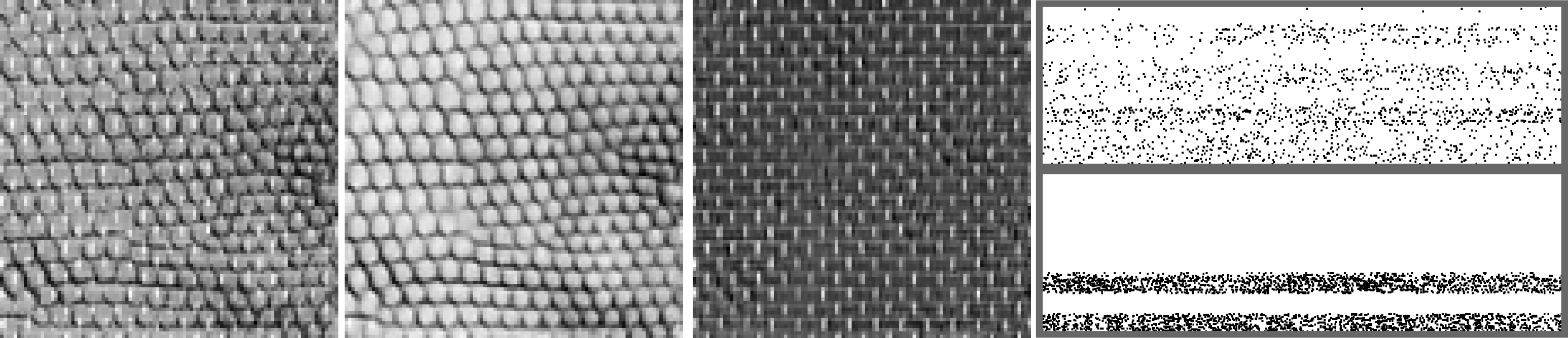}
\end{center}\vspace{-15pt}
\caption{\label{fig:texture}. {\footnotesize Results for the texture segmentation.  One example of the mixture and the C-HiLasso separated textures are shown. This is followed by the active set diagram (as in Figure \ref{fig:missing}), Lasso on top (with class selection wrongly all over the 8 textures) and C-HiLasso on bottom, where only the 2 corresponding groups are selected.}}
\end{figure}

We also used the digits dataset to experiment with missing data. We randomly discarded an average of 60\% of the pixels per mixed image
and then applied the C-Hilasso. The algorithm is capable of correctly detect which digits are present in the images. In Figure~\ref{fig:missing} we show some examples.
Note that this is a quite different problem than the
one commonly addressed in the matrix completion literature.
Here we do not aim to recover signals that all belong to a
unique unknown sub-space,
but signals that are the combination of two non-unique spaces
to be automatically selected from the available dictionary. Such unknown
spaces have common models/groups for all the signals in question
(the coarse level of the hierarchy), but not necessarily
the exact same atoms and therefore not necessarily belong to the same
sub-spaces. Both levels of the hierarchy are automatically detected,
e.g., that the
groups are those corresponding to ``3'' and ``5,'' and the exact
atoms (sub-spaces) in each group, these last ones possibly different for
each signal in the set.
While we consider that the possible sub-spaces are to be selected
from the provided dictionary, in Section~\ref{sec.discussion} we discuss learning
such dictionaries as well. In such case, the standard matrix
completion problem becomes a particular case of the C-HiLasso framework
(with a single group and all the signals having the same active set, sub-space,
in the group),
naturally opening numerous theoretical questions for this new more general
model.\footnote{Prof. Carin and collaborators have new results on the case of
a single group and signals in possible different sub-spaces of the group,
an intermediate model between standard matrix completion and C-HiLasso (personal communication).}
Finally, we used C-HiLasso to separate overlapping textures in an image. We chose 8 textures form
the Brodatz dataset and trained one dictionary for each one of them (these form the 8 groups of the dictionary). Then we created an image as the sum
of two textues (the testing images were not used in the training stage). In Figure~\ref{fig:texture} we show results.
The overall group Hamming distance obtained for C-HiLasso is 0.003, showing that the correct groups, and only them, were practically selected all the time.

\vspace{-5pt}
\section{Discussion} 
\label{sec.discussion}
In this paper we have introduced a new framework of collaborative hierarchical sparse coding, where multiple signals collaborate in their encoding, sharing code groups (models) and having (possible disjoint) sparse representations inside the corresponding groups. An efficient optimization approach was developed, which guarantees convergence to the global minimum, and examples illustrating the power of this framework were presented.
At the practical level, we are currently working on the applications of this proposed framework in a number of directions, including collaborative instruments separation in music; and signal classification, following the demonstrated capability to collectively select the correct groups/models.

At the theoretical level, a whole family of new problems is opened by this proposed framework. A critical one is the overall capability of selecting the correct groups and thereby of performing correct model selection and source separation. Let us consider for example the case of only two groups (so no sparsity at the group level) and a single signal composed by the linear combination of atoms from each group. Then, it is easy to show that the cross-mutual coherence between the groups plays a critical role. Let us call $\mu_i$, $i=1,2$, the internal coherence of the atoms of the group $i$, and $\mu_{1,2}$ the one between the groups (maximal normalized correlation between an atom of group $1$ with an atom of group $2$). Then it is easy to show that uniqueness of the separation can be guaranteed if $(2k_1-1)\mu_1+2k\mu_{1,2}<1$ and $(2k_2-1)\mu_2+2k_1\mu_{1,2}<1$, with $k_i$ the respective sparsity levels inside each group (this is a weaker bound that the more stringent one developed by \cite{Starck04imagedecomposition}).

This needs to be extended to actual sparsity at the group level and to the collaborative case. Note of course that considering a single active group is a particular case of our model (see \cite{CVPR} for works in this case), thereby an overall theoretical framework for our proposed collaborative hierarchical framework will automatically include numerous of the existing results in sparse coding.

Finally, we have also developed a framework for learning the dictionary for collaborative hierarchical sparse coding, meaning the optimization is simultaneously on the dictionary and the code. As it is the case with standard dictionary learning, this is expected to lead to significant performance improvements (again, see \cite{CVPR} for the particular case of this with a single group active at a time).


\noindent
{\bf Acknowledgments:} Work partially supported by NSF, ONR, NGA, and ARO. We thank Dr. Tristan Nguyen, when we presented him this model, he motivated us to think in a hierarchical fashion and to look at this as just the particular case of a fully hierarchical sparse coding framework. We also thank Prof. Larry Carin, Dr. Guoshen Yu, and Alexey Castrodad for very stimulating conversations and for the fact that their own work also motivated the example with missing information.



\end{document}